\documentclass[prl,twocolumn,showpacs,floatfix,superscriptaddress]{revtex4}

\usepackage{graphicx}

\def\ap#1,#2,#3#4{           {\it Ann. Phys. (NY)\/ }{\bf #1} #2 (19#3#4)}
\def\apj#1,#2,#3#4{          {\it Astrophys. J.\/ }{\bf #1} #2 (19#3#4)}
\def\apjl#1,#2,#3#4{         {\it Astrophys. J. Lett.\/ }{\bf #1} #2 (19#3#4)}
\def\app#1,#2,#3#4{          {\it Acta Phys. Polon.\/ }{\bf #1} #2 (19#3#4)}
\def\com#1,#2,#3#4{          {\it Comm. Math. Phys.\/ }{\bf #1} #2 (19#3#4)}
\def\ib#1,#2,#3#4{           {\it ibid.\/ }{\bf #1} #2 (19#3#4)}
\def\nat#1,#2,#3#4{          {\it Nature (London)\/ }{\bf #1} #2 (19#3#4)}
\def\np#1,#2,#3#4{           {\it Nucl. Phys.\/ }{\bf B#1} #2 (19#3#4)}
\def\npps#1,#2,#3#4{         {\it Nucl. Phys. B (Proc. Suppl.)\/ }{\bf B#1}
                             #2 (19#3#4)}
\def\plb#1,#2,#3#4{          {\it Phys. Lett.\/ }{\bf B#1} #2 (19#3#4)}
\def\pla#1,#2,#3#4{          {\it Phys. Lett.\/ }{\bf A#1} #2 (19#3#4)}
\def\prd#1,#2,#3#4{          {\it Phys. Rev.\/ }{\bf D#1} #2 (19#3#4)}
\def\prep#1,#2,#3#4{         {\it Phys. Rep.\/ }{\bf #1} #2 (19#3#4)}
\def\prl#1,#2,#3#4{          {\it Phys. Rev. Lett.\/ }{\bf #1} #2 (19#3#4)}
\def\pro#1,#2,#3#4{          {\it Prog. Theor. Phys.\/ }{\bf #1} #2 (19#3#4)}
\def\rmp#1,#2,#3#4{          {\it Rev. Mod. Phys.\/ }{\bf #1} #2 (19#3#4)}
\def\sp#1,#2,#3#4{           {\it Sov. Phys.-Usp.\/ }{\bf #1} #2 (19#3#4)}

\def\zp#1,#2,#3#4{           {\it Zeit. fur Physik\/ }{\bf C#1} #2 (19#3#4)}
\def\vecmet{\mbox{$\vec{\not\!\!{E}}_T$}}

\def\herw{{\sc  herwig}}

\def\pyt{{\sc  pythia}}

\def\ttb{\mbox{$t\bar{t}$}}

\def\ppb{\mbox{$p\bar{p}$}}

\def\vereq#1#2{\lower3pt\vbox{\baselineskip1.5pt \lineskip1.5pt}}

\def\beq{\begin{equation}}

\newcommand{\et}{\mbox{$E_{T}$}}
\newcommand{\pt}{\mbox{$P_{T}$}}

\newcommand{\bspace}{\!\!\!\!}
\newcommand{\met}{\mbox{$E{\bspace}/_{T}$}}
\newcommand{\niso}{\mbox{$N_{trk}^{iso}$}}

\newcommand{\ppbar}{\mbox{$p\overline{p}$}}

\def\eeq{\end{equation}}
\def\bea{\begin{eqnarray}}
\def\beaa{\begin{eqnarray*}}
\def\eea{\end{eqnarray}}
\def\eeaa{\end{eqnarray*}}
\def\bq{\begin{quote}}
\def\eq{\end{quote}}

\def\gappeq{\mathrel{\rlap {\raise.5ex\hbox{$>$}}
{\lower.5ex\hbox{$\sim$}}}}
\def\lappeq{\mathrel{\rlap{\raise.5ex\hbox{$<$}}
{\lower.5ex\hbox{$\sim$}}}}
\def\sm{standard model} % requested by A. Castro

\begin{document}
\hspace*{4.35in} \mbox{FERMILAB-Pub-03/285-E}
\bibliographystyle{revtex}
\title[lala]{Search for Kaluza-Klein Graviton Emission in $p\bar{p}$ 
Collisions \\at $\sqrt{s}=1.8$~TeV using the Missing Energy Signature \\
\vglue 0.15in
The CDF Collaboration\\
\vglue 0.15in}

%%%%%%%%%%%%%%%%%%%%%%%%%%%%%%%%%
\font\eightit=cmti8
\def\r#1{\ignorespaces $^{#1}$}
\hfilneg
\author{
\noindent
D.~Acosta,\r {14} T.~Affolder,\r 7 H.~Akimoto,\r {51}
M.G.~Albrow,\r {13} D.~Ambrose,\r {37}   
D.~Amidei,\r {28} K.~Anikeev,\r {27} J.~Antos,\r 1 
G.~Apollinari,\r {13} T.~Arisawa,\r {51}
% ...
}
\affiliation{
\r 1  {\eightit Institute of Physics, Academia Sinica, Taipei, Taiwan 11529, 
Republic of China} \\
\r 2  {\eightit Argonne National Laboratory, Argonne, Illinois 60439} \\
\r 3  {\eightit Istituto Nazionale di Fisica Nucleare, University of Bologna,
I-40127 Bologna, Italy} \\
\r 4  {\eightit Brandeis University, Waltham, Massachusetts 02254} \\
\r 5  {\eightit University of California at Davis, Davis, California  95616} \\
\r 6  {\eightit University of California at Los Angeles, Los 
Angeles, California  90024} \\ 
\r 7  {\eightit University of California at Santa Barbara, Santa Barbara, California 
93106} \\ 
\r 8 {\eightit Instituto de Fisica de Cantabria, CSIC-University of Cantabria, 
39005 Santander, Spain} \\
\r 9  {\eightit Carnegie Mellon University, Pittsburgh, Pennsylvania  15213} \\
\r {10} {\eightit Enrico Fermi Institute, University of Chicago, Chicago, 
Illinois 60637} \\
\r {11}  {\eightit Joint Institute for Nuclear Research, RU-141980 Dubna, Russia}
\\
\r {12} {\eightit Duke University, Durham, North Carolina  27708} \\
\r {13} {\eightit Fermi National Accelerator Laboratory, Batavia, Illinois 
60510} \\
\r {14} {\eightit University of Florida, Gainesville, Florida  32611} \\
\r {15} {\eightit Laboratori Nazionali di Frascati, Istituto Nazionale di Fisica
               Nucleare, I-00044 Frascati, Italy} \\
\r {16} {\eightit University of Geneva, CH-1211 Geneva 4, Switzerland} \\
\r {17} {\eightit Glasgow University, Glasgow G12 8QQ, United Kingdom}\\
\r {18} {\eightit Harvard University, Cambridge, Massachusetts 02138} \\
\r {19} {\eightit Hiroshima University, Higashi-Hiroshima 724, Japan} \\
\r {20} {\eightit University of Illinois, Urbana, Illinois 61801} \\
\r {21} {\eightit The Johns Hopkins University, Baltimore, Maryland 21218} \\
\r {22} {\eightit Institut f\"{u}r Experimentelle Kernphysik, 
Universit\"{a}t Karlsruhe, 76128 Karlsruhe, Germany} \\
\r {23} {\eightit High Energy Accelerator Research Organization (KEK), Tsukuba, 
Ibaraki 305, Japan} \\
\r {24} {\eightit Center for High Energy Physics: Kyungpook National
University, Taegu 702-701; Seoul National University, Seoul 151-742; and
SungKyunKwan University, Suwon 440-746; Korea} \\
\r {25} {\eightit Ernest Orlando Lawrence Berkeley National Laboratory, 
Berkeley, California 94720} \\
\r {26} {\eightit University College London, London WC1E 6BT, United Kingdom} \\
\r {27} {\eightit Massachusetts Institute of Technology, Cambridge,
Massachusetts  02139} \\   
\r {28} {\eightit University of Michigan, Ann Arbor, Michigan 48109} \\
\r {29} {\eightit Michigan State University, East Lansing, Michigan  48824} \\
\r {30} {\eightit Institution for Theoretical and Experimental Physics, ITEP,
Moscow 117259, Russia} \\
\r {31} {\eightit University of New Mexico, Albuquerque, New Mexico 87131} \\
\r {32} {\eightit Northwestern University, Evanston, Illinois  60208} \\
\r {33} {\eightit The Ohio State University, Columbus, Ohio  43210} \\
\r {34} {\eightit Osaka City University, Osaka 588, Japan} \\
\r {35} {\eightit University of Oxford, Oxford OX1 3RH, United Kingdom} \\
\r {36} {\eightit Universita di Padova, Istituto Nazionale di Fisica 
          Nucleare, Sezione di Padova, I-35131 Padova, Italy} \\
\r {37} {\eightit University of Pennsylvania, Philadelphia, 
        Pennsylvania 19104} \\   
\r {38} {\eightit Istituto Nazionale di Fisica Nucleare, University and Scuola
               Normale Superiore of Pisa, I-56100 Pisa, Italy} \\
\r {39} {\eightit University of Pittsburgh, Pittsburgh, Pennsylvania 15260} \\
\r {40} {\eightit Purdue University, West Lafayette, Indiana 47907} \\
\r {41} {\eightit University of Rochester, Rochester, New York 14627} \\
\r {42} {\eightit Rockefeller University, New York, New York 10021} \\
\r {43} {\eightit Instituto Nazionale de Fisica Nucleare, Sezione di Roma,
University di Roma I, ``La Sapienza," I-00185 Roma, Italy}\\
\r {44} {\eightit Rutgers University, Piscataway, New Jersey 08855} \\
\r {45} {\eightit Texas A\&M University, College Station, Texas 77843} \\
\r {46} {\eightit Texas Tech University, Lubbock, Texas 79409} \\
\r {47} {\eightit Institute of Particle Physics, University of Toronto, Toronto
M5S 1A7, Canada} \\
\r {48} {\eightit Istituto Nazionale di Fisica Nucleare, University of Trieste/\
Udine, Italy} \\
\r {49} {\eightit University of Tsukuba, Tsukuba, Ibaraki 305, Japan} \\
\r {50} {\eightit Tufts University, Medford, Massachusetts 02155} \\
\r {51} {\eightit Waseda University, Tokyo 169, Japan} \\
\r {52} {\eightit University of Wisconsin, Madison, Wisconsin 53706} \\
\r {53} {\eightit Yale University, New Haven, Connecticut 06520}}
\begin{abstract}
We report on  a search for direct Kaluza-Klein graviton production in
a data sample of \mbox{84 $\rm{pb}^{-1}$} of \ppb~collisions at
$\sqrt{\mbox{s}}$ = 1.8~TeV, recorded by the Collider Detector at
Fermilab.
We investigate the final state of large missing transverse
energy and one or two high energy jets. We compare the data with the
predictions from a $3+1+n$-dimensional Kaluza-Klein scenario in which 
gravity becomes strong at the TeV scale.
At  95\% confidence level (C.L.) for $n$~=~2, 4, and 6 we exclude an 
effective Planck scale below 1.0, 0.77, and 0.71~TeV, respectively.

\end{abstract}

\pacs{14.80.-j, 12.10.-g, 13.85.Rm}

\date{Sept 15 2003}

\maketitle

Early attempts to unify gravity and electromagnetism led to the idea 
of an extra circular spatial dimension~\cite{gn:14}.
Because of the periodicity of the extra dimension, 
the metric field of the five-dimensional spacetime is 
Fourier expandable in 
the extra dimension with 
four-dimensional fields (called Kaluza-Klein (KK) modes) as coefficients.  

More recently, Kaluza-Klein theories appear in scenarios of large extra 
dimensions as introduced by Arkani-Hamed, Dimopoulos, and Dvali 
(ADD)~\cite{ar:98}.  In these theories the \sm\ gauge theory is 
confined to a 3-dimensional domain wall (brane), embedded in a higher 
dimensional compactified bulk space.  Only gravity propagates in the full 
bulk space.  The $n$ compactified extra dimensions are assumed for simplicity 
to be ``large'' circles of common circumference $R$ (an $n$-torus).  As a 
result of compactification, the gravitational field that propagates in the 
bulk can be expanded in a series of states known collectively as the graviton KK 
tower.  Similar to a particle in a box, the momentum of the bulk field is 
quantized in the compactified dimensions.  For an observer trapped on the brane, 
each quantum of momentum in the compactified volume appears as a KK 
excited state with mass $m^2={\vec p_n}$$^{2}$, where ${\vec p_n}$ is
the momentum in the compactified dimensions, and with identical spin and gauge 
numbers. 

In such a model the Planck scale $M_{Pl}$, the radius $R$ of the compactified
space (here assumed to be a torus), 
and the new effective Planck scale $M_D$ are related by~\cite{Giudice:1999ck}
$$        M_{Pl}^2 = 8\pi R^n M^{2+n}_{D},      $$
where $n$ is the number of extra dimensions.
If $M_D$ takes values as low as a few TeV 
the Higgs naturalness problem~\cite{weinberg:78}
of the \sm\ can be solved by introducing a cutoff not too far above 
the electroweak scale with new physics 
entering at energies above this cutoff.  The hierarchy problem of the 
\sm\ is also recast: the question of why $M_{Pl}$ is so large 
compared to the $Z$ boson mass ($M_Z$) is replaced with the question of why $R$ is so large 
compared to $1/M_Z$, 
and an ultraviolet hierarchy problem is replaced with an infrared 
one. If we take the most optimistic case 
 of $M_{D} = 1$~TeV 
and use $M_{Pl}\sim 10^{19}$~GeV, 
we find that for $n = 1$, 2, 4 and 6, 
$R \sim 10^{11}$~m, 1~mm, 10~nm and 10~fm, respectively. 
 
All the states in the KK graviton tower, including the massless state, couple 
in an identical manner with universal strength of $M_{Pl}^{-1}$.  However 
there are $(ER)^n$ massive KK modes that are kinematically 
accessible in a collider process with energy $E$.  
The sum over the contribution from each KK state  
removes the Planck scale suppression 
and replaces it by powers 
of the fundamental scale $M_D\sim$~TeV.  The interactions of the massive 
KK graviton modes can then be observed in collider experiments 
either through their direct production and emission or through their virtual 
exchange in \sm\ processes~\cite{JHMS:02}. 

There are three processes in $\ppbar$ collisions that can result in
the emission of a graviton and a hadronic jet: $q\overline{q} \rightarrow gG$,
$qg \rightarrow qG$, and $gg \rightarrow gG$,
where $q$ and $g$ are quarks and gluons and $G$ is the graviton.  
The lowest-order Feynman diagrams for these 
processes are shown in Figure~1.  
%%%%%%%%%%%%%%%%%%%%%%%%%%%%%%%%%Figure 1%%%%%%%%%%%%%%%%%%%%%%%%%%%%%
\begin{figure}
\label{fig1}
\includegraphics[width=3.2in]{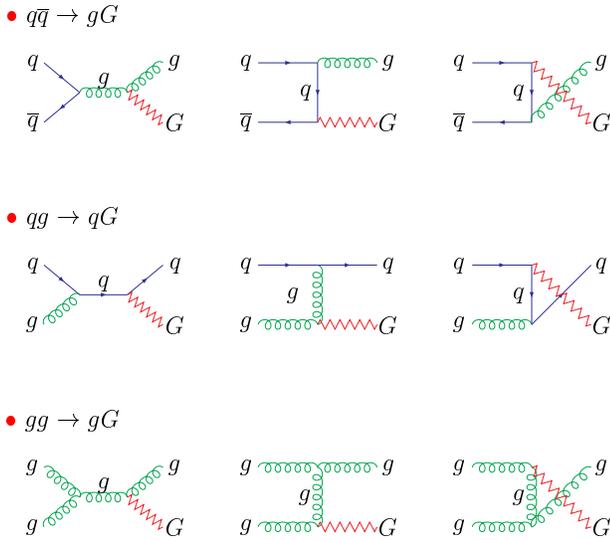}
\caption{Lowest-order Feynman diagrams for the emission of real gravitons in
           $p\overline{p}$ collisions.}
%           $q\overline{q} \rightarrow gG$,
%           $qg \rightarrow qG$, and
%           $gg \rightarrow gG$.}
\end{figure} 
%%%%%%%%%%%%%%%%%%%%%%%%%%%%%%%%%%%%%%%%%%%%%%%%%%%%%%%%%%%%%%%%%%%%%%%%%
The calculation of graviton emission is based on the effective low-energy 
theory that is valid below the scale $M_D$.  The corresponding Feynman 
rules are cataloged in~\cite{Han:1999sg,Giudice:1999ck}.
Since the graviton passes through the detector without decaying or interacting,
the experimental signature is missing transverse energy ($\met$)
from the emitted graviton and a hadronic jet
from the outgoing quark or gluon.

In this Letter we report the results of a search for the direct production 
of KK graviton modes using the rate of events with one or two energetic 
jets and large \met\ at the Collider Detector 
at Fermilab (CDF).  The search is based on \mbox{$84 \pm 4$ pb$^{-1}$} of
integrated luminosity recorded with the CDF detector during the 1994-95 
Tevatron run.

The CDF detector is described in detail in~\cite{cdf_d}.
The momenta of charged particles are measured in the central tracking 
chamber (CTC), which is positioned inside a 1.4~T superconducting 
solenoidal magnet.  Outside the magnet, electromagnetic and
hadronic calorimeters arranged in a projective tower geometry cover the
pseudorapidity region \mbox{$|\eta| < 4.2$~\cite{CDFcoo}} and are used
to identify jets. Jets are defined as localized energy depositions 
in the calorimeters and are reconstructed 
using an iterative clustering algorithm with a fixed cone of radius
\mbox{$\rm{\it{\Delta R}} \equiv \sqrt{\rm{\it{\Delta}}\eta^2 + \rm{\it{\Delta}}\phi^2} = 0.7$} in 
$\eta - \phi$ space~\cite{jets}. 
The transverse 
energy of a jet is $E_T=E\sin\theta$, where $E$ is the scalar sum of energy 
deposited in the calorimeter towers within the cone, and $\theta$ is
the angle formed by the beam-line and the cone axis~\cite{multiple_vertices}.
For this analysis, jets are required to have $E_T \geq 15$~GeV.

The missing transverse energy is defined as the negative vector sum of the
transverse energy in the electromagnetic and hadronic calorimeters,
\mbox{$\vecmet = -\sum_{i} (E_{i}\sin\theta_{i})\hat{n}_{i}$}, where $E_{i}$
is the energy of the {\it i}-th tower, $\hat{n}_{i}$ is a transverse
unit vector pointing to the center of each tower, and $\theta_{i}$ is
the polar angle of the tower.  The sum extends to $|\eta|=3.6$. 
The data sample was selected with an online trigger that 
requires $\met \equiv |\vecmet| > 30$~GeV. This is a sample dominated 
by instrumental backgrounds and by multijet events, 
where the observed missing energy is largely a result of
jet mismeasurements and detector resolution. 

The two-stage preselection we use to reject beam and
detector-related backgrounds, beam halo, and cosmic ray events
is described in~\cite{gluino_prl}. 
Events that pass the preselection are then required to have only
one or two jets with $E_T\ge 15$~GeV, with at least one jet
within $|\eta|< 1.1$. 

We remove events where the missing energy is due to energy flow 
from a jet to an
 uninstrumented region of the detector by requiring that 
the second highest \et\ jet does not point in $\eta$ to a detector gap
if it is within 0.5~radians in $\phi$ of the \met\ direction.
We reduce the residual  mismeasured multijet backgrounds
by requiring that the minimum $\delta\phi$ between the \met\ 
vector and any jet in the event ($\delta\phi_{min}$) is  greater than
0.3~radians 
and the $z$ position of the 
event vertex is within 60~cm of the nominal interaction point. 

To reduce the physics background contribution from 
electroweak processes with leptons in the final state
(dominated by $W(\rightarrow\ell\nu)$) 
%and $Z(\rightarrow\ell^+\ell^-)$  
we require that the two 
highest energy jets are not purely electromagnetic 
(by requiring the electromagnetic 
fraction $f_{em}\equiv E_{em}/E_{Tot}\le 0.9$) and the isolated track
multiplicity, \niso~\cite{niso} is zero.
For the final sample, we require $\met \geq 80$~GeV, $E_T \geq 80$~GeV for the leading jet and
$E_T \geq 30$~GeV for the second jet if there is more than 
one jet in the event. By accepting events with an energetic second jet we can reliably
normalize the background predictions from QCD simulation  
using the jet data, control the systematic uncertainty on the signal due to 
initial/final state radiation (ISR/FSR),
and interpret the results with a $K$-factor 
(the ratio of the cross sections at leading-order (LO) and
next-to-leading-order (NLO), $K=\sigma_{NLO}/\sigma_{LO}$) 
included in the estimated signal cross section.

The selection requirements and the number of events passing at each stage
are summarized in Table~1.
%~\ref{tab1}.

\begin{table}
\label{analysis}
\caption[The data selection path for the \met~plus one or two jets search]
        {The data selection path for the \met~plus one or two jets search.}
%\begin{ruledtabular}
\begin{tabular}{l|c}
\hline\hline 
{Selection Requirement}&{Events Passing}\\ \hline\hline
{Pre-Selection}&{300945}\\ \hline 
\multicolumn{1}{l}{$1 \le$ N$_{jet}\le 2$ (cone 0.7, $\et \ge 15$~GeV)}&
\multicolumn{1}{|c}{} \\
\multicolumn{1}{l}{$|\eta|(1~or~2) < 1.1$}&
\multicolumn{1}{|c}{157035} \\ \hline
\multicolumn{1}{l}{2nd jet gap veto}&
\multicolumn{1}{|c}{}\\ 
\multicolumn{1}{l}{$\delta\phi_{min}\ge 0.3$}&
\multicolumn{1}{|c}{} \\
\multicolumn{1}{l}{$|z_{vertex}|\le 60$~cm}&
\multicolumn{1}{|c}{50938}\\ \hline
\multicolumn{1}{l}{$f_{em}(1),~f_{em}(2)\le$ 0.9}&
\multicolumn{1}{|c}{21012} \\ \hline
\multicolumn{1}{l}{\niso=0}&
\multicolumn{1}{|c}{16459}\\ \hline
\multicolumn{1}{l}{$\et(1) \ge 80$~GeV}&
\multicolumn{1}{|c}{} \\
\multicolumn{1}{l}{If N$_{jet}=2$, $\et(2)\ge 30$~GeV} &
\multicolumn{1}{|c}{897} \\ \hline
{\met$\ge$80 GeV} &{284} \\ 
\hline\hline 
\end{tabular}
%\end{ruledtabular}
\end{table}

%%%%%%%%%%%%%%%%%%%%%%%%%%%%%%%%%%%%%%%%%%%%%%%%%%%%%%%%%%%%%%%%%%%%%%%%%%%%
Background events with missing energy and one or two jets arise from
\sm\ sources, predominantly $Z(\rightarrow \nu\bar{\nu})$+jets, 
$W(\rightarrow\ell\nu)$+jets $(\ell=\tau ,\mu ,e)$, 
and residual QCD production. 
While $Z(\rightarrow \nu\bar{\nu})$+jets produces real $\met$+jets,
$W(\rightarrow\ell\nu)$+jets mimics our signal when the lepton is
lost or misidentified.  To estimate the $Z$+jets and $W$+jets background 
levels and their uncertainties in the final sample we normalize 
\pyt~\cite{pythia}  Monte Carlo (MC) predictions using the observed 
$Z (\to e^+e^-)+$ jets data sample.  
QCD dijet events mimic our signal when one jet is badly mis-measured,
resulting in large $\met$.
For the QCD predictions we use the 
\herw\ MC program~\cite{herwig} and normalize to the high statistics 
jet data samples using well-balanced dijet events.  
We estimate additional backgrounds from \ttb, single 
top and diboson production using MC predictions~\cite{herwig,pythia}, which 
we normalize using the respective theoretical cross section calculations 
for these processes~\cite{tdib}.  

%%%%%%%%%%%%%%%%%%%%%%%%%%%%%%%%%%%%%%%%%%%%%%%%%%%%%%%%%%%%%%%%%%%%%%%%%%%%%%
The predicted backgrounds from \sm\ processes are summarized in Table~2.
Of the 274 total events predicted to pass our selection requirements, 
160 are predicted to come from $Z(\rightarrow \nu\bar{\nu})$+jets,
89 from the combined $W(\rightarrow\ell\nu)$+jets
electroweak processes, and 22 from QCD production.
Because the MC predictions have been normalized to high statistics data
samples, the dominant uncertainty on the $W+$jets and $Z+$jets predictions
is the 4\% luminosity uncertainty.  The QCD prediction has an additional
14\% uncertainty due to jet energy resolution~\cite{gluino_prl}.
We observe 284 events in the data.  
In Figure~2 the predicted \sm\ \met\ distribution is compared with the 
distribution we observe in the data.  
In Figure~3 the same comparison is shown for other kinematic distributions. 
In both figures the data are consistent with the expected background.
An additional contribution from graviton production would result in a
smooth excess over the background 
in nearly all the kinematic distributions, as shown in Figure~4 for \met . 
%%%%%%%%%%%%%%%%%%%%%%%%%%%%%%Table 2%%%%%%%%%%%%%%%%%%%%%%%%%%%%%%%%%%%%%%%
\begin{table}
\label{bkgds}
\caption{The predicted number of events in the final sample from 
\sm\ sources and the number observed in the data.}
%\begin{ruledtabular}
\begin{center}
\begin{tabular}{l|c} \hline\hline
{Background Source}          & {Predicted Events}\\\hline\hline
{$Z(\to\nu \bar\nu)$+jets}   & {160.2 $\pm$ 11.5} \\\hline
{$W(\to\tau\nu)$+jets}       & {46.6 $\pm$ 5.5} \\\hline
{$W(\to\mu\nu)$+jets}        & {23.8 $\pm$ 5.0} \\\hline
{$W(\to e\nu)$+jets}         & {18.1 $\pm$ 4.3} \\\hline
{QCD}                        & {21.7 $\pm$ 6.7} \\\hline
{\ttb, single $t$, dibosons} & { 3.9 $\pm$ 0.3} \\\hline\hline
{Total predicted} & {274.1 $\pm$ 15.9} \\\hline\hline
{Observed}        & {284} \\
\hline\hline
\end{tabular}
\end{center}
%\end{ruledtabular}
\end{table}
%%%%%%%%%%%%%%%%%%%%%%%%%%%%%%Table 2%%%%%%%%%%%%%%%%%%%%%%%%%%%%%%%%%%%%%%%

%%%%%%%%%%%%%%%%%%%%%%%%%%%%%%Figure 2%%%%%%%%%%%%%%%%%%%%%%%%%%%%%%%%%%%%%%%
\begin{figure}
\label{met_comp}
\includegraphics[width=3.2in]{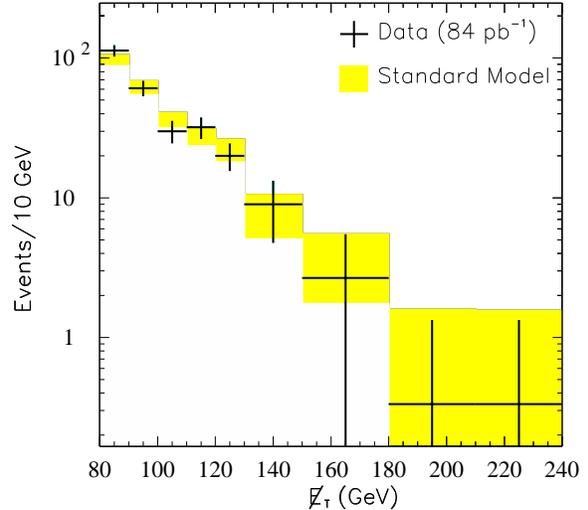}
\caption{ Comparison between data (points) and \sm\ predictions (boxes) 
of the~\met\ distribution.  There are 284 data events, to be compared 
with 274$\pm$16 events predicted from \sm\ sources.  The distribution 
is plotted with a variable bin size; the contents for bin sizes 
greater than 10~GeV are normalized accordingly. The height of the boxes 
shows the uncertainty on the \sm\ prediction.}
\end{figure}
%%%%%%%%%%%%%%%%%%%%%%%%%%%%%%Figure 2%%%%%%%%%%%%%%%%%%%%%%%%%%%%%%%%%%%%%%%

We use the \pyt\ MC program  to generate 
datasets of graviton emission, using the leading-order production cross sections
calculated in~\cite{Giudice:1999ck}.  The signal processes are 
simulated for $n=2$, 4, and 6 extra dimensions, and for a range of 
values of $M_D$.
The signal efficiency ranges from 2.9\% for two extra dimensions to 
6.4\% for six extra dimensions (due to different relative weights
 of the three production processes of Figure~1) and is largely independent of $M_D$.
The total relative systematic uncertainty on the signal efficiency
is 25\%, mostly due to modelling of ISR/FSR  
(21\%), 
jet energy scale (11\%), renormalization scale (8\%) 
and parton density functions (2\%)~\cite{mrsd-:93}. 

%%%%%%%%%%%%%%%%%%%%%%%%%%%%%%Figure 3%%%%%%%%%%%%%%%%%%%%%%%%%%%%%%%%%%%%%%%
\begin{figure}
\label{other_dist}
\includegraphics[width=3.2in]{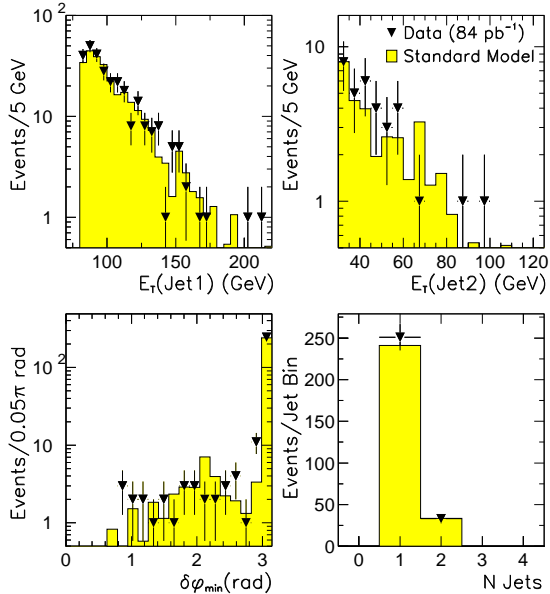}
\caption{Comparison between data (points) and \sm\ predictions (histogram)
of the first and second leading jet \et, $\delta\phi_{min}$, and $N_{jet}$
distributions.}
\end{figure}
%%%%%%%%%%%%%%%%%%%%%%%%%%%%%%%%%%%%%%%%%%%%%%%%%%%%%%%%%%%%%%%%%%%%%%%%%%

%%%%%%%%%%%%%%%%%%%%%%%%%%%%%%Figure 4%%%%%%%%%%%%%%%%%%%%%%%%%%%%%%%%%%%%%%%
\begin{figure}
\label{met_referees}
\includegraphics[width=3.2in]{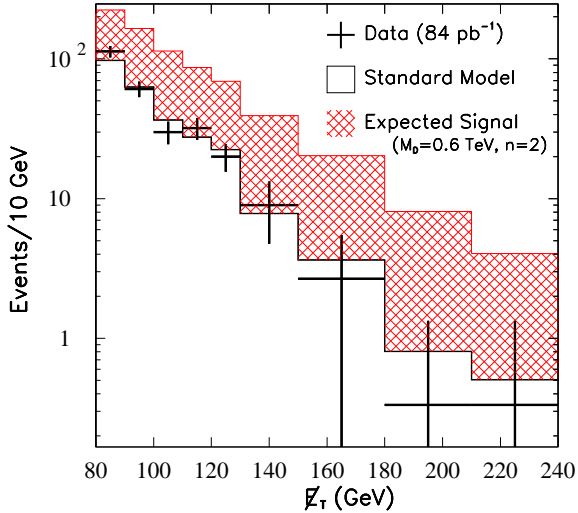}
\caption{The predicted~\met\ distribution from \sm\ processes (histogram) and 
the one from the expected graviton signal (for $n=2$, $M_{D}$ = 0.6~TeV, and  
a $K$-factor of 1.0) added to the standard model (hatched).  
The signal appears as a smooth excess over the \sm\ background. The points are 
the observed data.}
\end{figure}
%%%%%%%%%%%%%%%%%%%%%%%%%%%%%%Figure 4%%%%%%%%%%%%%%%%%%%%%%%%%%%%%%%%%%%%%%%

Using a Monte Carlo technique to convolute the uncertainty on the
background estimate with the relative systematic uncertainty on
the signal efficiency, the 95\% C.L.~\cite{lim} upper limit on the 
number of signal events is 62.  As shown in Figure~5, for $K=1.0$ we 
exclude an effective Planck scale less than 1.00~TeV for $n=2$, 
less than 0.77~TeV for $n=4$, and less than 0.71~TeV for $n=6$.  
Recently the D\/0 collaboration 
reported a limit on direct graviton emission using a $K$-factor of 1.3 
in the signal cross section~\cite{referee1}. They report limits 
of 0.99~TeV for $n=2$, 0.73~TeV for $n=4$
and 0.65~TeV for $n=6$. For direct comparison, using $K=1.3$ our corresponding lower 
limits on $M_D$ are 1.06~TeV for $n=2$, 0.80~TeV for $n=4$, and 0.73~TeV for $n=6$.

%%%%%%%%%%%%%%%%%%%%%%%%%%%%%%Figure 5 %%%%%%%%%%%%%%%%%%%%%%%%%%%%%%%%%%%%%%%
\begin{figure}
\label{exp_evt}
\includegraphics[width=3.2in]{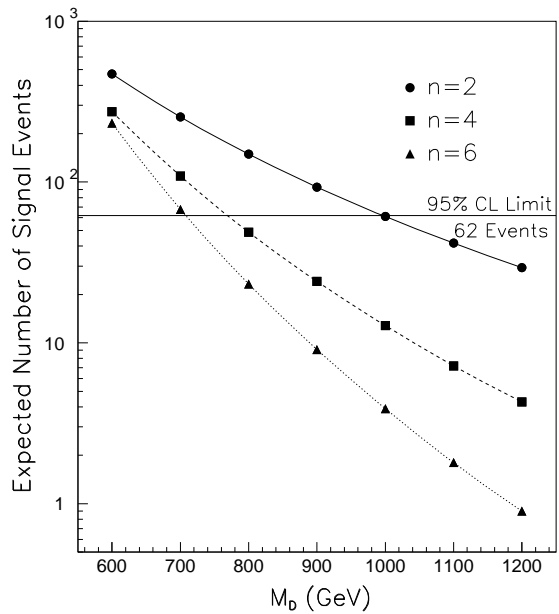}
\caption{The three curves show the number of expected signal events for 
$n=2$, 4, and 6 extra dimensions as a function of the effective Planck 
scale $M_D$ for a $K$-factor of 1.0.  The straight line shows the 95\% 
C.L. upper limit on the number of signal events.}
\end{figure}
%%%%%%%%%%%%%%%%%%%%%%%%%%%%%%Figure 3 %%%%%%%%%%%%%%%%%%%%%%%%%%%%%%%%%%%%%%%
These are the best limits to date on direct graviton emission~\cite{referee1,other} 
from the Tevatron.

Assuming compactification on a torus, 
the limits on $M_D$ with $K=1.0$ correspond to limits on the compactification
radius of $R<0.48$~mm for $n=2$, $R<0.014$~nm for $n=4$, and 
$R<42$~fm for $n=6$.

We thank the Fermilab staff and the technical staffs of the
participating institutions for their vital contributions.  This work was
supported by the U.S. Department of Energy and National Science Foundation;
the Italian Istituto Nazionale di Fisica Nucleare; the Ministry of Education,
Science, Sports and Culture of Japan; the Natural Sciences and Engineering 
Research Council of Canada; the National Science Council of the Republic of 
China; the Swiss National Science Foundation; the A. P. Sloan Foundation; the
Bundesministerium fuer Bildung und Forschung, Germany; the Korea Science 
and Engineering Foundation (KoSEF); the Korea Research Foundation; the 
Comision Interministerial de Ciencia y Tecnologia, Spain; and the Pritzker Foundation. 
We thank Konstantin Matchev and Joe Lykken for providing the 
MC generation code for graviton production in $p\bar{p}$  
collisions~\cite{newpythia}.

%\begin{acknowledgments}
%\end{acknowledgments}
%\bibliography{your bib file}
%\bibliography{biblio.tex}

\begin{thebibliography}{99}
%0
\bibitem{gn:14} G. Nordstrom, 
Z. Phys. {\bf 15}, 504 (1914); T. Kaluza, Preuss. Akad. Wiss, Berlin, Math. Phys. 
K {\bf 1}, 966 (1921); O. Klein, Z. Phys. {\bf 37},
895; Nature {\bf 118}, 516 (1926). 
%1
\bibitem{ar:98} N.~Arkani-Hamed, S.~Dimopoulos and G.~R.~Dvali,
Phys.\ Lett.\ B {\bf 429}, 263 (1998).

\bibitem{Giudice:1999ck}
G.~F.~Giudice, R.~Rattazzi and J.~D.~Wells,
%``Quantum gravity and extra dimensions at high-energy colliders,''
Nucl.\ Phys.\ B {\bf 544}, 3 (1999).
%%CITATION = HEP-PH 9811291;%%
%

\bibitem{weinberg:78} Steven Weinberg Phys.\ Lett.\ B {\bf 82}, 387 (1979); 
Eldad Gildener Phys. \ Rev. \ D {\bf 14}, 1667 (1976).

%4
\bibitem{JHMS:02}E.~A.~Mirabelli, M.~Perelstein and M.~E.~Peskin, Phys. Rev. Lett. {\bf 82}, 2236 (1999).  For a review, see J.~Hewett and M.~Spiropulu,
 Ann. Rev. Nucl. Part. Sci., Vol. {\bf 52}: 397-424 (2002).

%3
\bibitem{Han:1999sg}
T.~Han, J.~D.~Lykken and R.~J.~Zhang,
%``On Kaluza-Klein states from large extra dimensions,''
Phys.\ Rev.\ D {\bf 59}, 105006 (1999)
[arXiv:hep-ph/9811350].
%%CITATION = HEP-PH 9811350;%%
%%

\bibitem{cdf_d}
F. Abe {\it et al.}, Nucl. Inst. and Methods,  
{\bf A271}, 387 (1988). 

%8
\bibitem{CDFcoo} 
In the CDF coordinate system, $\phi$ and $\theta$ are the
azimuthal and polar angles with respect to
the proton beam direction.
The pseudorapidity $\eta$
is defined as ${\rm ln}[\tan (\theta/2)]$.

%9
\bibitem{jets}
 F. Abe {\it et al.}, CDF Collaboration,
Phys. Rev. {\bf D45}, 1448 (1992).

%10
\bibitem{multiple_vertices} If there are multiple vertices in the
event we use the vertex with the largest $\sum P_{T}$ of associated tracks.

\bibitem{gluino_prl}T. Affolder {\it et al.}, CDF Collaboration,
Phys. Rev. Lett. {\bf 88}, 041801 (2002).
% M. Spiropulu, Ph.D Thesis, Harvard University (2000).

%14
\bibitem{niso}
\niso  is the number of high momentum isolated tracks in the event.
Tracks qualify as such if  they have 
transverse momentum \pt~$\ge$~10~GeV/$c$, 
impact parameter $d_{0}\le$~0.5~cm, 
vertex difference $|z_{track}-z_{event}|< 5$~cm  
and the total transverse momentum  $\Sigma\pt$ 
of all tracks (with impact parameter $d_{0}\le 1$~cm) 
around them in a cone of 
$\rm{\it{\Delta R}} \equiv \sqrt{\rm{\it{\Delta}}\eta^2 + \rm{\it{\Delta}}\phi^2} = 0.4$ 
is $\Sigma\pt \le 2$~GeV/$c$. 

%5
\bibitem{pythia}
T. Sj\"{o}strand, Comput. Phys. Commun. {\bf 82}, 74 (1994).
\pyt\ {\sc v6.115} is used.
%6

\bibitem{herwig}
G. Marchesini {\it et al.}, Comput. Phys. Commun. {\bf 67}, 465 (1992).
\herw\, v5.6 is used. See hep-ph/9607393 (1996). 
%18
\bibitem{tdib} R.~Bonciani {\it et al.}, 
Nucl. Phys. {\bf B529}, 424 (1998); 
T.~Stelzer, Z.~Sullivan and S.~Willenbrock,
Phys. Rev. {\bf D54}, 6696 (1996);
 T.~Tait and C.~P.~Yuan, hep-ph/9710372 (1997);
J.~Ohnemus {\it et al.}, Phys. Rev. {\bf D43}, 3626 (1991); 
Phys. Rev.  {\bf D44}, 1403 and 3477 (1991);
T. Affolder {\it et al.}, Phys. Rev. {\bf D64} 032002 (2001).

%18
%19
\bibitem{mrsd-:93}MRSD- is used as our reference structure function; A.~D.~Martin, W.~J.~Stirling and R.~G.~Roberts,
%{\it New information on parton distributions},
Phys.\ Rev.\  {\bf D47}, 867, (1993).
%%CITATION = PHRVA,D47,867;%%

\bibitem{lim} G. Zech, Nucl. Instrum. Methods {\bf A277}, 608 (1989); T.Huber {\it et al.}, Phys Rev {\bf D41}, 2709 (1990).

\bibitem{referee1} V.~M.~Abizov {\it et al.}, D\/0 collaboration,
Phys. Rev. Lett. {\bf 90}, 251802 (2003).
  
%\bibitem{other} Recently the D\/0 collaboration (arXiv:hep-ex/0302014) 
\bibitem{other} 
D.~Acosta {\it et al.}, CDF Collaboration, 
  Phys. Rev. Lett. {\bf 89}, 281801 (2002).
For a summary of LEP results on graviton emission, 
see G.~Landsberg, {\it ``Extra Dimensions and More,''}
[arXiv:hep-ex/0105039].
%%CITATION = HEP-EX 0105039;
%% See also, P.Abreu {\it et al.}, The DELPHI Collaboration, Eur. Phys. J. C {\bf 17}, 53 (2000); 
%G. Abbiendi {\it et al.}, The OPAL Collaboration, Eur. Phys. J. C {\bf 18}, 253 (2000).
%7
%
% ****** End of file template.aps ******
\bibitem{newpythia}  The current version of \pyt(6.2) includes these processes.
T.~Sj\"ostrand, P.~Ed\'en, C.~Friberg, L.~L\"onnblad, G.~Miu, S.~Mrenna and E.~Norrbin, 
Comput. Phys. Commun. {\bf 135}, 238 (2001).
%arXiv:hep-ph/0108264.

\end{thebibliography}
\par

\end{document}